\documentclass[useAMS,usenatbib]{mn2e}
 \usepackage{graphicx}
 \usepackage[cp1251]{inputenc}
\usepackage[T2A]{fontenc}

 \title[Search for the evolutionary relationship]
 {Search for the evolutionary relationship between Galactic globular and open clusters using data from the Gaia DR2 Catalog}

 \author[A. T. Bajkova and V. V. Bobylev]
        {A. T. Bajkova\thanks{E-mail: anisabajkova@mail.ru} and V. V. Bobylev\\
     Central (Pulkovo) Astronomical Observatory of RAS, 65/1
     Pulkovskoye Chaussee, Saint Petersburg, 196140, Russia}

\begin{document}
 \date{Received 2019 July 20; in original form 2019 April 27}
 \pagerange{\pageref{firstpage}--\pageref{lastpage}} \pubyear{2019}
 \maketitle
 \label{firstpage}

\begin{abstract}
Passing through the Galactic disk, a massive object such as a
globular cluster, can trigger star formation process leading to the birth of open clusters. Here, we analyze such possible evolutionary relationship between globular and open clusters. To search for the closest rapprochement between objects we computed backwards the orbits of 150 Galactic globular and 232 open clusters (younger than 100 Myr) with proper motions, derived from the Gaia DR2 Catalog. The orbits were computed using the recently modified  three-component (disk, bulge and halo)
axisymmetric Navarro-Frenk-White potential, which was complemented by non-axisymmetric bar and spiral density wave potentials. We obtained a new estimate for the frequency of  impacts of globular clusters about the Galactic disk, which is equal to 4 events for 1 million years. In the framework of the considered scenario, we highlight the following nine pairs of globular and open clusters, with rapprochement within 1 kpc at the time of the intersection the Galactic disk by a globular cluster for the latest 100 Myr:
 NGC~104 -- Turner~3,
 NGC~104 -- NGC~6396,
 NGC~104 -- Ruprecht~127,
 NGC~5139 -- Trumpler~17,
 NGC~5139 -- NGC~6520,
 NGC~6341 -- NGC~6613,
 NGC~6838 -- NGC~6520,
 NGC~7078 -- NGC~7063,
 NGC~6760 -- Ruprecht~127.

\end{abstract}


 \section{INTRODUCTION}
As it is known, the main characteristics of globular and open
star clusters in the Galaxy differ significantly. Globular clusters
(GCs) are distributed throughout the whole Galaxy, contain $10^4-10^6$ members,
which for a long time retain gravitational connectedness, on
average, for more than 10 Gyr. Open clusters (OCs) belong to the Galactic disk, contain no more than
$10^4$ members and can completely dissipate for
a few Gyr after their birth. Indeed, numerical simulations of the dynamical evolution of OCs, made by
\cite{Chumras06}, show stellar tails stretched along the
Galactic cluster orbits developing with time, which must completely dissolve
and mix with the stellar background in a time near 2 Gyr \citep{Kup08}.

GCs and OCs have different formation mechanisms. Details of GCs formation retain unclear. At the same time, there is the hypothesis that the most massive $\omega$~Cen-type GCs can be remnants of destroyed dwarf satellite galaxies of the Milky Way. The mechanism of formation of OCs has been worked out in more detail. Origination of an OC begins with the collapse of a part of a giant molecular cloud, a cold dense cloud of gas and dust. Numerous factors that violate the balance of a giant molecular cloud,  may result in the formation of an OC.  They are: shock waves from nearby supernovae, collisions with other clouds, tidal interactions with the bulge, disk and spiral arms. It seems also possible that the formation of OCs may be triggered by a GC passing through the Galactic disk.

The question of detection of observational effects of crossing the Galactic disk by globular clusters was, apparently for the first time, raised in \cite{Brosche91}. These authors, in particular, studied the motion of globular clusters NGC~362 and NGC~6218. In \cite{Rees-Cudworth03}, the pair of a globular cluster NGC~6397 and a young scattered stellar cluster NGC~6231 was considered.

The idea behind the evolutionary relationship between OCs and GCs implies that when a GC passes through the Galactic disk it may approach a giant molecular cloud so closely that the latter will be taken out of its state of equilibrium. That is, under favorable circumstances, the propagation of GCs through the disk can trigger the formation of OCs or stellar associations consisting of several OCs.

\cite{VandePutte09} analyzed Galactic trajectories of 54 globular clusters with measured proper motions,  radial velocities and distances. They found that potentially promising in the above sense are globular clusters NGC~3201, NGC~6397 and NGC~6838. Based on more extensive and reliable data, the analysis was repeated in \cite{BobBajk19AstBull} using a sample of 133 Galactic globular clusters, in which pairs  NGC~104 -- Ruprecht~129 and NGC~6362 -- Pismis~11 were recognized as the most interesting. And in \cite{BobBajk18},
the hypothesis of the formation of the Gould Belt after intersection of the disk by a globular cluster $\omega$~Cen was analyzed as well.

The second data release of the Gaia mission \citep{Lindegren18} has provided five astrometric parameters for more than 1.3 billion stars in the Milky Way. The mean errors of the trigonometric parallax and both proper motion components in this catalogue depend on the stellar magnitude. For example, the parallax errors lie within the range 0.02--0.04 mas for bright stars $(G<15^m)$ and are 0.7 mas for faint stars $(G=20^m).$ For more than 7 million stars of spectral types F--G--K, their radial velocities were determined with a mean error of $\sim$1 km s$^{-1}$.

The aim of this paper is to study the Galactic orbits of known globular clusters in order to search for a suitable candidate whose passage through the Galactic disk could have triggered the formation of open star clusters. To this end, we use the values of proper motions of globular and open clusters derived by various authors from Gaia DR2 Catalog. In addition, new values for OCs distances were calculated from trigonometric parallaxes of their likely members from the Gaia DR2 Catalog.

The paper is structured as follows. The data on GCs and OCs are described in Section 2. The method based on analyzing the orbits of GCs and OCs is considered in Section 3. The model for the axisymmetric Galactic potential is described in Section 3.1. The bar and Galactic spiral structure potentials are considered in Section 3.2. Equations of motion are given in Section 3.3. The statistical modeling principle is described in Section 3.4. The temporal characteristics of star formation are  given in Section 3.5. The results are presented and discussed in Section 5.

 \section{DATA}

In this paper, the main source of data on globular clusters is the \cite{Vasiliev19} Catalog. It contains average proper motions of 150 globular clusters calculated from the data of the Gaia DR2 Catalog.

We used \cite{Cantat-Gaudin18} as the main source of mean values of OCs proper motions and parallaxes, calculated from the data of the Gaia DR2 Catalog. These values were determined using the most likely members of clusters. The average values of heliocentric radial velocities of the OCs were taken from the
MWSC (Milky Way Star Clusters, \cite{MWSC13}) Catalog, and in some cases, from \cite{Soubiran18}, where they were calculated exclusively according to the Gaia DR2 Catalog. Estimates of the OCs ages were taken from the MWSC Catalog.

The authors of the Catalog Gaia DR2 \citep{Lindegren18} already noted the presence of a systematic shift in
Gaia DR2 parallaxes:  $\Delta\pi=-0.029 $ mas relative to the inertial coordinate system. The minus sign means that the correction must be added to the Gaia DR2 parallaxes to reduce them to the standard values.
Later in \cite{Stassun-Torres18}, \cite{Yalyalieva18}, \cite{Riess18} and \cite{Zinn18} it was shown that the value of this correction is underestimated. Based on these works, we added to all the original OCs parallaxes the correction $0.050$~mas.

As a result, we selected 232 OCs younger than 100 Myr ($\lg t<8 $) with known proper motions, parallaxes ($\sigma_\pi/\pi<30\% $), and radial velocities.

\section{METHOD}
Since the method used in our study is based on the analysis of the orbits of GCs and OCs, we will spend some time describe our model for the Galaxy gravitational potential. We believe that the most realistic
model available is our refinement \cite{Bajkova-Bob16}, \cite{Bajkova-Bob17} of the Navarro-Frenk-White (NFW) three-component (bulge, disk, halo) axisymmetric model \citep{NFW97}, supplemented with terms that take into account the influence of the central bar and the spiral density wave.

 \subsection{Model for the Axisymmetric Galactic Potential}
We present a model for the axisymmetric gravitational potential (ASGP) of the Galaxy as a sum of three components: the central spherical bulge, $\Phi_b(r(R,Z))$, the disk $\Phi_d(r(R,Z))$, and the massive, spherical dark-matter halo $\Phi_h(r(R,Z))$:
 \begin{equation}
 \begin{array}{lll}
  \Phi(R,Z)=\Phi_b(r(R,Z))+\Phi_d(r(R,Z))+\Phi_h(r(R,Z)).
 \label{pot}
 \end{array}
 \end{equation}
Here, we used a cylindrical coordinate system ($R,\psi,Z$) with
its origin at the Galactic center. In Cartesian coordinates
$(X,Y,Z)$ with their origin at the Galactic center, the distance
to a star (the spherical radius) is $r^2=X^2+Y^2+Z^2=R^2+Z^2,$
where the $X$ axis is directed from  the Galactic
center toward the Sun, the $Y$ axis is perpendicular to the $X$ axis and points
in the direction of the Galactic rotation, and the $Z$ axis is
perpendicular to the Galactic $(XY)$ plane and points in the
direction of the North Galactic pole. The gravitational potential
is expressed in units of 100 km$^2$ s$^{-2}$, distances in kpc,
masses in units of the mass of the Galaxy, $M_{gal}=2.325\times
10^7 M_\odot$, and the gravitational constant is taken to be
$G=1.$

 \begin{table}
 \begin{center}
 \caption[]
 {\small\baselineskip=1.0ex
 Parameters of the Galactic potential model, $M_0=2.325\times10^7 M_\odot$
  }
 \label{t:model-III}
 \begin{tabular}{|c|r|}\hline
 Parameter     &     Value         \\\hline
 $M_b$ ($M_0$) &    443     \\
 $M_d$ ($M_0$) &   2798     \\
 $M_h$ ($M_0$) &  12474   \\
 $b_b$ (kpc)   & 0.2672 \\
 $a_d$ (kpc)   &   4.40  \\
 $b_d$ (kpc)   & 0.3084 \\
 $a_h$ (kpc)   &    7.7    \\\hline
 \end{tabular}
  \end{center}
  \end{table}

The potentials of the bulge $\Phi_b(r(R,Z))$ and disk $\Phi_d(r(R,Z))$ were taken to have the form proposed by \cite{Miyamoto-Nagai75}:
 \begin{equation}
  \Phi_b(r)=-\frac{M_b}{(r^2+b_b^2)^{1/2}},
  \label{bulge}
 \end{equation}
 \begin{equation}
 \Phi_d(R,Z)=-\frac{M_d}{\Biggl[R^2+\Bigl(a_d+\sqrt{Z^2+b_d^2}\Bigr)^2\Biggr]^{1/2}},
 \label{disk}
\end{equation}
where $M_b$ and $M_d$ are the masses of the corresponding
components and $b_b, a_d,$ and $b_d$ are scale parameters of the
components in kpc. The halo component was taken in accordance with
\cite{NFW97}:
 \begin{equation}
  \Phi_h(r)=-\frac{M_h}{r} \ln {\Biggl(1+\frac{r}{a_h}\Biggr)}.
 \label{halo-III}
 \end{equation}
Table \ref{t:model-III} presents the parameters of the model for the Galactic potential (\ref{bulge})--(\ref{halo-III}) from \cite{Bajkova-Bob16}, \cite{Bajkova-Bob17}, computed using the rotational velocities of Galactic objects at distances $R$ out to $\sim$200 kpc, which were obtained by \cite{Bhat14} with $R_\odot=8.3$ kpc for the Galactocentric distance and $V_\odot=244$ km s$^{-1}$ for the linear velocity of the local standard of rest around the Galactic center. When deriving the corresponding Galactic rotation curves, we adopted the mentioned values for $R_\odot$ and $V_\odot$.

 \subsection{The Bar and The Spiral Density Wave Potentials}

The terms describing the potential of the bar and the spiral density wave are added to the right-hand side of formula (\ref{pot}), in general forming non-axisymmetric gravitational potential (NASGP) of the Galaxy.

We adopted the bar potential in the form of the triaxial ellipsoid model in accordance with \cite{Paloush93}:

\begin{equation}
  \Phi_B = -\frac{M_B}{(q_B^2+X^2+[Y\cdot a_B/b_B]^2+[Z\cdot a_B/c_B]^2)^{1/2}},
\label{bar}
\end{equation}
where $M_B$ is the mass of the bar, which is equal to $43.1\times$ M$_{gal}$; $a_B, b_B,$ and $c_B$ are the three semiaxes of the bar $(a_B/b_B=2.381, a_B/c_B=3.03);$ $q_B$ is the length of the bar;
$X=R\cos\vartheta$ and $Y=R\sin\vartheta$, where $\vartheta=\theta-\Omega_B \cdot t-\theta_B$, $\theta$ is the initial position angle of the object: $\tan \theta=Y_0/X_0$ ($X_0,Y_0$ are the initial coordinates of the object in the Cartesian coordinate system in accordance with (\ref{init}), $\Omega_{B}$ is the circular velocity of the bar, $t$ is time, $\theta_B$ is the bar orientation angle relative to Galactic axes $X,Y$, which is counted from the line connecting the Sun and the Galactic center (the $X$-axis) to the major axis of the bar in the direction of Galactic rotation. We adopted the estimates of the bar parameters $\Omega_B=55$~km s$^{-1}$ kpc$^{-1}$, $q_B=8$~kpc, and $\theta_B=45^{\circ}$ from \cite{BobBajk16}.

When the spiral density wave is taken into account \citep{Lin-Shu64}, the right-hand side of the formula (\ref{pot}) is supplemented with the term \citep{Fernandez08}:
\begin{equation}
 \Phi_{sp} (R,\theta,t)= A\cos[m(\Omega_p t-\theta)+\chi(R)],
 \label{Potent-spir}
\end{equation}
where
 $$
 A= \frac{(R_\odot\Omega_\odot)^2 f_{r0} \tan i}{m}
 $$
 and
 $$
 \chi(R)=- \frac{m}{\tan i} \ln\biggl(\frac{R}{R_\odot}\biggr)+\chi_\odot.
 $$
Here, $A$ is the amplitude of the spiral-wave potential, $\Omega_\odot$ the angular velocity of the Galaxy
at the solar distance $R_\odot$, $f_{r0}$
the ratio of the radial component of the perturbation to the total
gravitation of the Galaxy, $\Omega_p$ the angular velocity of the
wave's rigid-body rotation, $m$ the number of spiral arms, $i$ the
pitch angle of the arms ($i<0$ for a trailing pattern), $\chi$ the
phase of the radial wave ($\chi=0^\circ$ corresponds to the center
of the arm), and $\chi_\odot$ the Sun's radial phase in the spiral
wave. We adopted the following parameters for the spiral wave:
 \begin{equation}
 \begin{array}{lll}
 i=-13^\circ,\\
 f_{r0}=0.05,\\
 \chi_\odot=-120^\circ,\\
 \Omega_p=20~\hbox {km s$^{-1}$ kpc$^{-1}$}
 \label{param-spiral}
 \end{array}
 \end{equation}
for the four-armed model of the pattern $m = 4$. The `` -- '' sign
of $\chi_\odot $ means that we are counting this angle from the
Carina--Sagittarius arm.

 \subsection{Equations of Motion}
The equations of motion of a test particle in the Galactic potential have the form
\begin{equation}
 \begin{array}{llllll}
 \dot{X}=p_X, ~~\dot{Y}=p_Y, ~~\dot{Z}=p_Z,\\
 \dot{p}_X=-\partial\Phi/\partial X,\\
 \dot{p}_Y=-\partial\Phi/\partial Y,\\
 \dot{p}_Z=-\partial\Phi/\partial Z,
 \label{eq-motion}
 \end{array}
\end{equation}
where $p_X, p_Y,$ and $p_Z$ are canonical momenta, and a dot denotes a derivative with respect to time. We integrated Eqs.~(\ref{eq-motion}) using a fourth-order Runge-Kutta algorithm.

We took the peculiar velocity of the Sun relative to the Local Standard of Rest to be $(u_\odot,v_\odot,w_\odot)=(11.1,12.2,7.3)$~km s$^{-1}$, as was determined by \cite{Schonrich10}. Here, heliocentric velocities correspond to a set of moving Cartesian coordinates, with $u$ directed towards the Galactic center, $v$ in the direction of the Galactic rotation, and $w$ perpendicular to the Galactic plane,
towards the north Galactic pole.

Let the initial positions and space velocities of a test particle in the heliocentric coordinate system be
$(x_0,y_0,z_0,u_0,v_0,w_0)$. The initial positions and velocities of the test particle in Galactic Cartesian coordinates are then given by
\begin{equation}
 \begin{array}{llllll}
 X_0=R_\odot-x_o, Y_0=y_o, Z_0=z_o+h_\odot,\\
   U=-(u_0+u_\odot),\\
   V=v_0+v_\odot+V_\odot,\\
   W=w_0+w_\odot,
 \label{init}
 \end{array}
\end{equation}
where $h_\odot=16$~pc is the height of the Sun above the Galactic plane \citep{BobBaj16}.

 \begin{figure}
  \includegraphics[width=0.95\columnwidth]{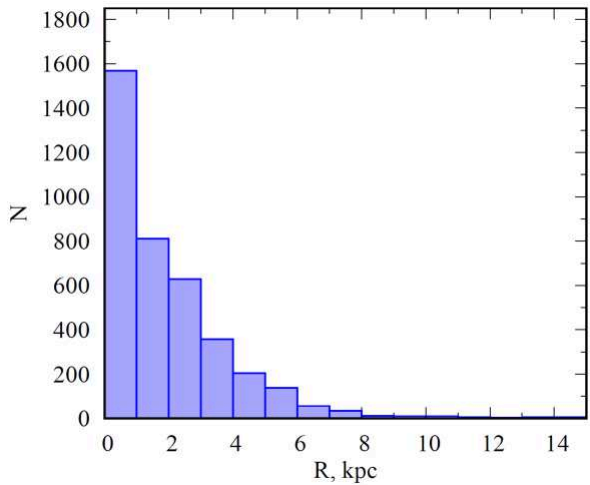}
  \caption{Histogram  of the number of the crossings of the
Galactic plane $XY$ by GCs during the latest billion years versus
the distance $R$ from the rotation axis of the Galaxy.
  }
  \label{f1}
 \end{figure}

 \subsection{Statistical Modeling}
To determine the confidence domain of the intersection points of the GC of the galaxy disk, as well as the OC location area at the moments of GC impacts, depending on the accuracy of the data, we performed Monte-Carlo statistical modeling for each GC--OC pair of interest.

In this simulation, before integration of the orbits, we add to the initial phase space $ (x_0, y_0, z_0,u_0, v_0, w_0)$ of the object the normally distributed random errors with zero means and standard deviations $(\sigma_{x_0}, \sigma_{y_0}, \sigma_{z_0}, \sigma_{u_0}, \sigma_{v_0}, \sigma_{w_0})$. The number of random realizations in each Monte-Carlo experiment was equal to 1000. It was empirically established that this number is optimal in terms of achievable accuracy and computational costs.

We determined the required standard deviations $(\sigma_{x_0}, \sigma_{y_0}, \sigma_{z_0}, \sigma_{u_0}, \sigma_{v_0}, \sigma_{w_0})$, also using the Monte-Carlo statistical simulation while calculating the initial coordinates and velocities of objects from the data on  positions, heliocentric distances, proper motions and radial velocities taking into account their normally distributed errors with zero means and standard deviations known from measurements.
 
Note that the confidence domains for the  points where the GCs cross the Galactic plane, as well as for the location of the corresponding OCs at the time of the GC crossing, are estimated at the 99.7\% $(3\sigma)$ probability level.

Obviously, the intersection of the confidence domains of GC and OC is the basis for determining the probability of the evolutionary relationship between GC and OC.

 \begin{table*} \caption[]
 {\small
 Parameters of GC--OC pairs with a possible evolutionary coupling according to Gaia DR2 data
   }
  \begin{center}  \label{t:02}  \small
  \begin{tabular}{|l|l|l|l|l|l|l|l|l|l|l|l|l|}
   \hline
 GC      &      OC    & $\Delta r_t,$ kpc& $\Delta r_t,$ kpc&$t,$ Myr backwards &$t_A,$ & $t-t_A,$ & $N_{cr}$\\
         &            & in ASGP  & in NASGP  & in ASP &Myr    &  Myr     &         \\\hline
 NGC 104 &   NGC 5606         & 1.16&1.37 &51.5& 22.4&29.1&I&\\
 NGC 104 &   NGC 6396         & 1.61&0.94 &51.5& 32.0&19.5&I&*\\
 NGC 104 &   Ruprecht 127     & 1.18&0.74 &51.5& 35.5&16.0&I&*\\
 NGC 104 &   Turner 3         & 0.73&0.63 &51.5& 28.8&22.7&I&*\\
 NGC 104 &   Lynga 6          & 1.52&1.62 &51.5& 28.2&23.3&I&\\
 NGC 104 &  vdBergh-Hagen 202 & 1.10&1.15 &51.5& 39.8&11.7&I&\\
    \hline
Palomar 13 &   NGC 433           &1.87 &1.83 &93.5 &64.6&28.9&I&\\
Palomar 13 &   NGC 7654          &1.51 &1.51 &93.5 &79.4&14.1&I&\\\hline

NGC 5139&  NGC 6520          &0.95 &0.67   &81.5 &56.9&24.6&II&*\\
NGC 5139&  Trumpler 17       &0.68 &0.65   &81.5 &59.6&21.9&II&*\\\hline

NGC 6341&  NGC 6613          &1.44&0.59  &61.5 &50.7&10.8&II&*\\
NGC 6341&  Ruprecht 127      &1.51&1.63  &61.5 &35.5&26.0&II&\\
NGC 6341&  Ruprecht 164      &1.29&1.16  &61.5 &50.1&11.4&II&\\
NGC 6341&  vdBergh-Hagen 202 &1.17&1.64  &61.5 &39.8&21.7&II&\\\hline

Palomar 7&   NGC 6910        &1.96&1.80 &53.5 &33.9&19.6&II&\\\hline

NGC 6838&  NGC 6520          &1.25&0.79   &71.5&56.9&14.6&II&*\\
NGC 6838&  NGC 6613          &1.18&0.63   &71.5&50.7&20.8&II&*\\
NGC 6838&  Ruprecht 164      &1.26&1.47   &71.5&50.1&21.4&II&\\
NGC 6838&  Trumpler 17       &0.99&1.09   &71.5&59.6&11.9&II&\\\hline

NGC 7078&  Basel 8          &1.49&0.68   &101.5&84.1&17.4&II&*\\
NGC 7078&  NGC 7654         &1.49&0.62   &101.5&79.4&22.1&II&*\\
NGC 7078&  Trumpler 2       &1.38&1.18   &101.5&84.1&17.4&II&\\
NGC 7078&  ASCC 113         &1.72&1.33   &101.5&85.1&16.4&II&\\
NGC 7078&  NGC 7063         &0.96&0.50   &101.5&90.2&11.3&II&*\\\hline

NGC 6760&  Ruprecht 127     &0.71&0.48   &64.5 &35.5&29.0&III&*\\
NGC 6760& vdBergh-Hagen 202 &1.85&1.32   &64.5 &39.8&24.7&III&\\\hline
 \end{tabular}\end{center} \end{table*}

 \begin{table*} \caption[]
 {\small
 Initial heliocentric space and velocity coordinates and their uncertainties for the GCs and OCs listed in f table \ref{t:02}}
  \begin{center}  \label{t:03}  \small
  \begin{tabular}{|c|l|c|c|c|c|c|c|c|c|c|c|c|c|}
   \hline
Object   & Name &$x_0$&$y_0$&$z_0$&$u_0$&$v_0$&$w_0$&$\sigma_{x_0}$&$\sigma_{y_0}$&$\sigma_{z_0}$&$\sigma_{u_0}$&$\sigma_{v_0}$&
$\sigma_{w_0}$  \\
   &             & kpc &kpc &kpc &km/s&km/s&km/s&kpc &kpc &kpc &km/s&km/s&km/s   \\\hline

  &NGC 104    &1.87& -2.58&  -3.18& -88.6&  -80.8&  37.9&  0.09&  0.12&  0.15& 3.9&  4.3&  1.2\\
  &NGC 5139   &3.16& -3.89&   1.34&  87.7& -269.5& -88.2&  0.15&  0.19&  0.06& 2.6&  4.7&  7.3\\
  &NGC 6341   &2.51&  6.33&   4.74& -36.8& -208.9&  87.0&  0.12&  0.31&  0.23& 0.2&  5.7&  7.7\\
  &NGC 6760   &5.96&  4.35&  -0.50&  73.2& -105.8& -21.8&  0.29&  0.21&  0.02& 3.8&  5.2&  1.1\\
GC&NGC 6838   &2.18&  3.32&  -0.31&  52.0&  -58.0&  31.7&  0.10&  0.16&  0.01& 3.1&  1.9&  1.4\\
  &NGC 7078   &3.90&  8.37&  -4.77&  89.7& -194.3& -35.5&  0.19&  0.42&  0.24& 6.5&  5.4&  4.2\\
  &Palomar 7  &4.98&  1.99&   0.53& 191.2&  -63.0&  20.1&  0.24&  0.09&  0.02& 2.4&  5.9&  0.3\\
  &Palomar 13 &0.96& 19.11& -17.66&-179.2&  -36.7& -87.4&  0.04&  0.94&  0.87&12.5&  5.9&  6.8\\\hline
  &NGC 433           & -1.14&  1.57&  -0.08&  31.9&  -22.8&  -6.1&   0.11&  0.16&  0.01& 1.2&  0.9&  1.3\\
  &NGC 5606          &  1.65& -1.66&   0.04& -71.2&  -16.9& -11.0&   0.15&  0.15&  0.01& 4.3&  4.4&  1.4\\
  &NGC 6396          &  2.47& -0.26&  -0.07& -31.6&  -20.0&  -7.5&   0.38&  0.04&  0.01& 2.4&  3.8&  1.9\\
  &Ruprecht 127      &  2.27& -0.28&  -0.10& -31.9&  -18.7&  -9.1&   0.24&  0.03&  0.01& 0.3&  2.6&  1.5\\
  &Ruprecht 164      &  1.60& -3.74&   0.04& 123.8&  -49.9& -16.4&   0.26&  0.60&  0.01&12.0& 11.1&  3.2\\
  &Turner 3          &  1.54&  0.33&  -0.03& -12.4&  -17.9& -11.6&   0.12&  0.03&  0.01& 9.7&  2.6&  1.4\\
  &Lynga 6           &  2.00& -1.13&   0.01& -69.0&   -1.2&  -9.2&   0.23&  0.13&  0.01& 2.2&  3.7&  1.6\\
  &vdBergh-Hagen 202 &  1.59& -0.44&   0.04& -56.7&  -17.7&  -8.6&   0.14&  0.04&  0.01& 1.3&  2.9&  1.1\\
OC&NGC 7654          & -0.60&  1.42&   0.01&  30.5&  -30.7&  -3.4&   0.04&  0.10&  0.01& 2.6&  4.9&  1.1\\
  &NGC 6520          &  1.65&  0.08&  -0.08& -22.5&   -1.3&  -5.1&   0.19&  0.01&  0.01& 0.9&  0.9&  1.2\\
  &Trumpler 17       &  0.75& -2.22&   0.01& -45.8&  -43.0&   0.4&   0.08&  0.23&  0.01& 6.7&  9.6&  0.9\\
  &NGC 6613          &  1.40&  0.35&  -0.02& -11.8&  -10.4&   0.3&   0.08&  0.02&  0.01& 4.3&  1.5&  0.7\\
  &NGC 6910          &  0.33&  1.64&   0.05&  43.2&  -41.9&  -3.2&   0.03&  0.14&  0.01& 4.6&  2.7&  1.2\\
  &Basel 8           & -1.35& -0.59&  -0.00&  -3.9&  -17.3&  -8.4&   0.07&  0.03&  0.01& 6.6&  3.1&  0.7\\
  &Trumpler 2        & -0.49&  0.45&  -0.04&  -4.0&  -11.8& -13.4&   0.02&  0.01&  0.01& 0.5&  0.6&  0.7\\
  &ASCC 113          &  0.06&  0.54&  -0.06&   4.9&   -6.0&  -7.5&   0.01&  0.01&  0.01& 0.6&  3.2&  0.5\\
  &NGC 7063          &  0.07&  0.64&  -0.11&   2.4&    2.5&  -9.4&   0.01&  0.01&  0.01& 0.8&  5.7&  1.1\\\hline

 \end{tabular}\end{center}
 \end{table*}

 \begin{table*} \caption[]
 {\small
 Parameters of 9 selected GC--OC pairs with the closest rapprochements ($\Delta r_t<1$ kpc) in the total gravitational potential accounting for bar and spiral density wave}
  \begin{center}  \label{t:04}  \small
  \begin{tabular}{|l|l|c|c|c|c|c|c|c|c|c|}
   \hline
 GC  & OC          & $\Delta r_t,$ &$t$  backwards, &$t_A,$& $t-t_A,$ &$\phi_{cr},$& $V_{cr},$& Mass of GC,    &Relative impact& $P$  \\
     &             &          kpc  & Myr            &  Myr &  Myr     &  $\deg $    & km/s    &  $10^6 M_\odot$ &energy of GC   &   \\\hline
 NGC 104&Turner 3      & 0.63 &51.5 &28.8   & 22.7   & 34 & 173 &0.78&0.0500   &0.51 \\
 NGC 104&NGC 6396      & 0.94 &51.5 &32.0   & 19.5   & 34 & 166 &0.78&0.0480   &0.33 \\
 NGC 104&Ruprecht 127  & 0.74 &51.5 &35.5   & 16.0   & 34 & 166 &0.78&0.0480   &0.33 \\
 NGC 5139&Trumpler 17  & 0.65 &83.5 &59.6   & 23.9   & 39 & 353 &3.55&1.0000   &0.33 \\
 NGC 5139&NGC 6520     & 0.67 &83.5 &56.9   & 26.6   & 39 & 319 &3.55&0.8200   &0.21 \\
 NGC 6341&NGC 6613     & 0.59 &62.5 &50.7   & 11.8   & 34 & 327 &0.33&0.0800   &0.51 \\
 NGC 6838&NGC 6520     & 0.79 &73.5 &56.9   & 16.6   & 12 & 73  &0.05&0.0006   &0.15 \\
 NGC 6760&Ruprecht 127 & 0.48 &64.5 &35.5   & 29.0   & 19 & 137 &0.25&0.0100   &0.33 \\
 NGC 7078&NGC 7063     & 0.50 &102.5&90.2   & 12.3   & 35 & 393 &0.80&0.2800   &0.30 \\\hline
 \end{tabular}\end{center}
 \end{table*}


 \subsection{Time Characteristics of Star Formation}
It is clear that some time must elapse after the impact of a GC on the Galactic plane before stars will be formed. Following~\cite{VandePutte09}, we based our study on the relation
\begin{equation}
 t=t_{\rm C}+t_{\rm SF}+t_{\rm A},
 \label{time-3}
\end{equation}
where $t$ is the time elapsed from the crossing of the Galactic
disk by the GC to the present time, $t_{\rm C}$ the time between
the crossing and the onset of star formation, $t_{\rm SF}$ the
duration of the star formation, and $t_{\rm A}$ the age of the
structure formed.

The value of the first term in (\ref{time-3}) is known only with a large
uncertainty, and is in the range 0--30 million years. For example,
$t_{\rm C}=15$~million years according to the estimate of
\cite{LepineDuvert94} obtained from simulations of an impact of a
high-velocity cloud onto the disk. According to \cite{Wallin96},
this time interval is $t_{\rm C}=30$~million years. In the model
computations of \cite{Bekki09}, the time interval for star
formation is in the range $t_{\rm C}=7-15$~million years.
According to \cite{McKee02}, the second term is $t_{\rm SF}=0.2$
million years (for a stellar mass $M>1 M_\odot)$; since this is
small compared to the other terms, we can neglect it in a first
rough estimate.

\begin{figure*}
{\begin{center}
   \includegraphics[width=0.90\textwidth]{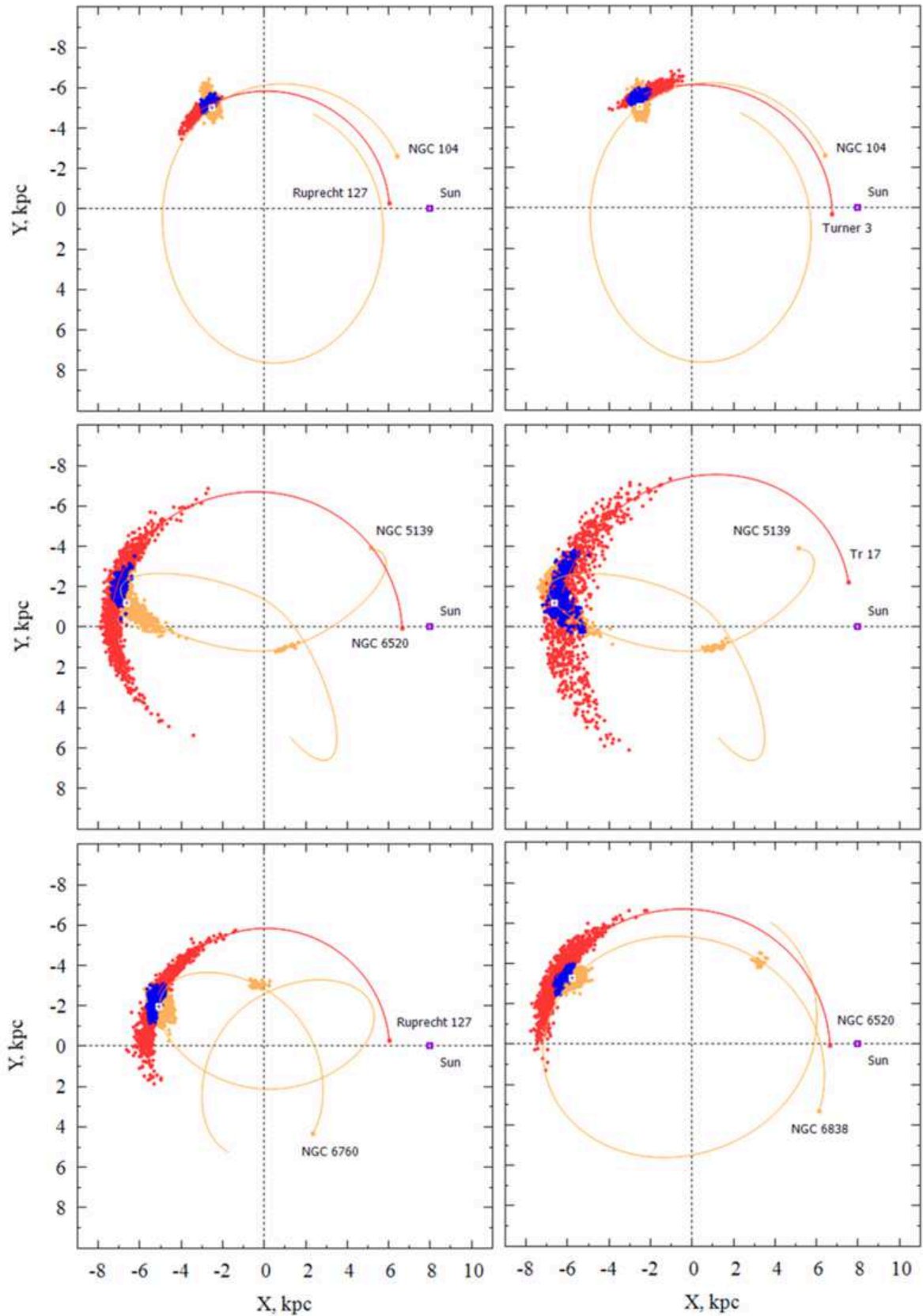}
 \caption{\small
The orbits and confidence domains in the $ XY $ Galactic plane for
pairs NGC~104 -- Ruprecht~127, NGC~104 -- Turner~3, NGC~5139 --
NGC~6520, NGC~5139 -- Trumpler~17, NGC~6760 -- Ruprecht~127,
NGC~6838 -- NGC~6520 from table \ref{t:02}. Explanation is given
in the text.
  }
  \label{f2}
\end{center}}
\end{figure*}
\begin{figure*}
{\begin{center}
   \includegraphics[width=0.90\textwidth]{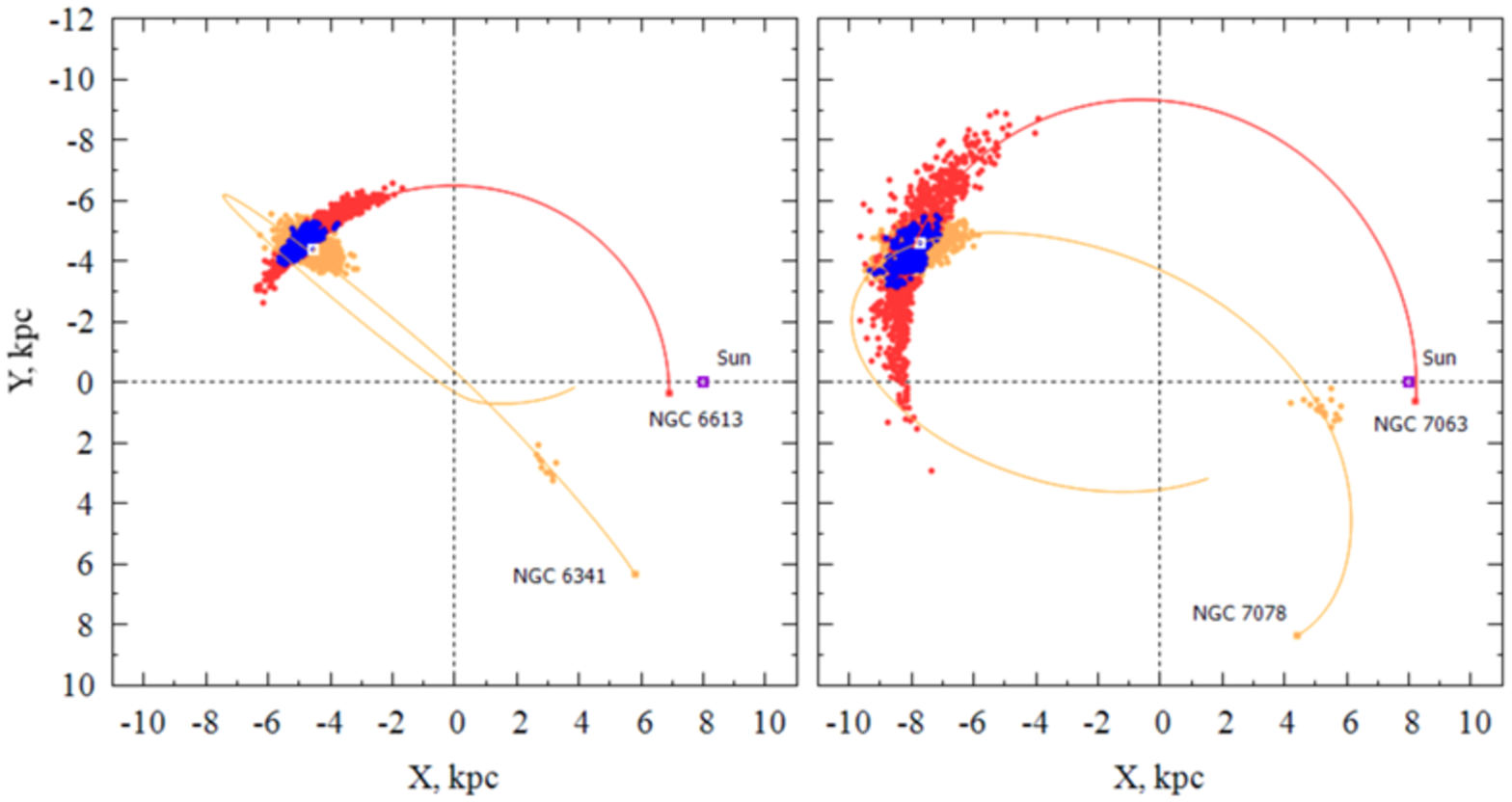}
 \caption{\small
The orbits and confidence domains in the $ XY $ Galactic plane for
pairs NGC~6341 -- NGC~6613, NGC~7078 -- NGC~7063 from table
\ref{t:02}. Explanation is given in the text.
  }
\label{f3}
\end{center}}
\end{figure*}

\section{RESULTS AND DISCUSSION}
Fig.~\ref{f1} gives the number of intersections of the Galactic plane by globular clusters from our entire sample during latest billion years versus the distance to the rotation axis of the Galaxy. Based on these data, we estimated the average number of intersections of the Galactic disk by globular clusters to be
equal to 4 events for a million years. According to works ~ \cite{Salerno09, VandePutte09}, this value is 1 event for a million years. In \cite{BobBajk19AstBull}, an estimate equal to 3 events within a million years,  which is closer to the present one, was obtained. As it is seen from Fig.~\ref{f1}, most of the impacts occur in the central region of the Galactic disk.

Next, a search was carried out for new couples, potential  GC -- OC  pairs, with possible evolutionary relationship. The search was based on calculation of closest rapprochement of GCs and OCs orbits in the past. The Galactic orbits of the objects were built both in the axisymmetric modified NFW potential and the potential, which includes the central bar and the spiral density wave components.

The search results are presented in table~\ref{t:02}.

The initial heliocentric coordinates $(x_0,y_0,z_0)$ and velocities $(u_0,v_0,w_0)$, needed for calculation of orbits of the objects listed in table \ref{t:02}, are given in columns 3--8 of table \ref{t:03}. The coordinate and velocity uncertainties: standard deviations $(\sigma_{x_0},\sigma_{y_0},\sigma_{z_0})$ and $(\sigma_{u_0},\sigma_{v_0},\sigma_{w_0})$, are given in columns 9--14 of table \ref{t:03}. These standard deviations were used in the Monte-Carlo statistical modeling as it is described in Section 3.4.

The first and second columns of table \ref{t:02} present the designations of GCs and OCs; the third and fourth columns, the values of rapprochement $\Delta r_t$, obtained in axisymmetric and non-axisymmetric potentials respectively;  the fifth column, the moment $ t$ of crossing the Galactic disc by GC, calculated using axisymmetric potential; the sixth column, the age of the OCs $ t_A$;  the seventh column, the difference $ t-t_A $; and the last one , the serial number $N_{cr}$ of the intersection (the last one is designated as "I"; the penultimate, "II"; the third from the last, "III" ). Note that the table contains only the pairs with $\Delta r_t\leq2$~kpc for which $10<t-t_A <30$~Myr. In addition, the obvious condition $t_A <t$ must be satisfied in accordance with the relation (\ref{time-3}).

As can be seen from the table~\ref{t:02}, virtually for every GC there are several suitable OC candidates. This is consistent with the results of \cite{VandePutte09}, where for the first time
it was shown that the formation of associations consisting of multiple OCs is possible. \cite{BobBajk19AstBull} also noted such possibility. However, the names of specific clusters in these works mostly disagree, excepting only the globular clusters NGC~104 and NGC~5139 ($\omega$~Cen) mentioned by all authors.

Comparison of rapprochements $\Delta r_t$ obtained in axisymmetric potential and the potential accounting for bar and spiral structure (ASGP and NASGP) shows the degree of the influence of the non-axisymmetric potential components.

As it can be seen from table \ref{t:02}, the same open cluster may have an evolutionary relationship with different globular clusters. This means that star formation in giant molecular clouds can be caused by the successive effects of two or several globular clusters crossing the Galactic disk at different but sufficiently close to each other moments of time. Thus, the formation of OC Rupprecht~127 can have been caused by the impacts of two globular clusters, NGC~6760 and NGC~104, that occurred 71 and 51 million years ago respectively. A similar situation may have taken place  with OCs NGC~6520 and NGC~6613. Open cluster vdBergh-Hagen~202 may have been born as a result of the impacts of three GCs: NGC~6760, NGC~6341 and NGC~104, 64, 61 and 51 million years ago respectively.

Based on the typical size of a giant molecular cloud of 700–1000 pc, we have identified pairs (designated by asterisk in the last column of table \ref{t:02}), which have $\Delta r_t\leq1$ kpc in full gravitational potential.

Table~\ref{t:04} presents parameters of 9 selected GC--OC pairs with the closest rapprochement. Here, except of $\Delta r_t $, $t_A$ and time intervals $ t, t-t_A $, five more parameters are given. These are: the angle $\phi_{cr}$ at which a GC crosses the Galactic plane; the velocity $ V_{cr}$ that a GC has relative to
the environment at time of crossing; mass of a GC; impact energy of a GC relative to the highest energy of GC NGC~5139 in pair with OC Trumpler~17; probability $P$ of the relationship between corresponding GC and OC.
Probability $P$ was determined using Monte-Carlo experiment as it is described in Section 3.4.

Based on numerical simulation of an intersection of the Galactic disk by a massive ($3.3\times10^6M_\odot $) high-velocity cloud, \cite{Comeron92}, \cite{Comeron94} have shown that star formation will be the most effective in case of oblique falling object. Therefore, we are  interested in the angle $ \phi_{cr}. $ As we can see from table~\ref{t:04}, all the obtained angles are acute.

For the pairs listed in table~\ref{t:04}, Figs.~\ref{f2} and \ref{f3} are presented. Since OCs Ruprecht~127 and NGC~6396 have very close orbits (compare their initial parameters and ages from tables \ref{t:03} and \ref{t:02}), we provide graphical illustration only for the pair NGC~104 -- Ruprecht~127. As it can be seen from table \ref{t:04}, almost all parameters of these two pairs are equal. This fact makes it possible to assume that the formation of both open clusters is very likely to have been caused by the same reason, namely, the intersection of the disk of the Galaxy by the NGC~104 51 million years ago.

In the Figures, we display the Galactic orbits of the globular and open clusters, calculated using the total
gravitational potential consisting of three component (bulge, disk, halo) axisymmetric potential and non-axisymmetric potentials of bar and spiral density wave. The orbits of the GCs (drawn by gray, red in color version) are integrated on the interval of 200 million years backwards. The OCs orbits (drawn by light, yellow in color version) are integrated on the time interval $t$ up to the moment of intersection of the
Galactic plane by a corresponding globular cluster. The points in the Figures are obtained using
Monte-Carlo simulations and make up the confidence areas corresponding to the level $ 3\sigma $. The brightest points (yellow in color) correspond to the orbit end points of OCs. Gray dots (red) fill the confidence region of the intersection of the Galaxy disk by GCs. The darkest points (blue) indicate the intersection of the areas corresponding to OCs and GCs. Since we used 1000 model iterations, the ratio of the number of points in the intersection area to 1000 yields the probability $P$. As it can be seen from table \ref{t:04}, pairs NGC~104 -- Turner~3 and NGC~6341 -- NGC~6613 have the highest probability equal to 0.51.

The masses of most GCs are known with an accuracy of 15--20 \%. In table \ref{t:04}, masses of our GCs are listed according to the work \cite{Baum-Hilker18}. Given the velocity of $ V_{cr}, $ which a GC has at the time of the impact relative to the supposed hydrogen cloud, we can conclude that  NGC~5139 ($\omega$~Cen) had the highest impact energy (see table \ref{t:04}). The second place belongs to NGC~7078, then go NGC~6341 and NGC~104. The pair NGC~6838 -- NGC~6520 shows both the lowest impact energy and the smallest probability of  evolutionary relationship.

\section*{Conclusions}
The passage of globular clusters through the giant molecular clouds of the Galactic disk can initiate a star formation process leading to the generation of open star clusters and even entire associations. Our study, devoted to the establishment of links between the passage of GCs through the Galactic disk and the formation of OCs, is based on the search for past rapprochements between the GCs and OCs with taking into account the age of the OCs and the duration of the star formation process. Since the search for the rapprochements involves the integration of orbits in the Galactic gravitational potential, the accuracy of data and the model of the Galactic gravitational potential, are of great importance. In this paper, an attempt to solve this problem on the basis of data from the Gaia DR2 catalog about 150 GCs and 232 young OCs has been made. As a model of the gravitational potential, we chose the three-component (disk, bulge and halo) axisymmetric Navarro-Frenk-White potential model, which we recently updated from the most up-to-date data on circular velocities of various objects in a wide range of galactocentric distances \citep{Bajkova-Bob16}. The study was carried out both in the axisymmetric potential and in the potential that takes into account the Galactic bar and the spiral density wave with parameters that meet modern estimates.

The following results of our research can be distinguished:

1.A new estimate of the frequency of GCs impacts on the Galactic disk was obtained:
 4 events for 1 million years. It is shown that most of these impacts occurred in the central region of the Galactic disk.

2.The parameters of nine GC –- OC pairs  with the closest rapprochement within 1 kpc at the time of the intersection the Galactic disk by a globular cluster for the latest 100 Myr were obtained.  These are the following pairs:

 NGC~104 -- Turner~3,

 NGC~104 -- NGC~6396,

 NGC~104 -- Ruprecht~127,

 NGC~5139 -- Trumpler~17,

 NGC~5139 -- NGC~6520,

 NGC~6341 -- NGC~6613,

 NGC~6838 -- NGC~6520,

 NGC~7078 -- NGC~7063,

 NGC~6760 -- Ruprecht~127.

For these pairs, a probability of possible evolutionary relationship was estimated using Monte-Carlo simulation. The probability is maximal, equal to 0.51 for pairs NGC~104 -- Turner~3 and NGC~6341 -- NGC~6613. Globular cluster NGC~5139 ensures the maximal impact energy in the pairs with open clusters Trumpler~17 and NGC~6520.

 \section*{Acknowledgments}
The authors are grateful to the anonymous referee for critical remarks that contributed to an improvement of the paper. The authors are thankful to Kirill Maslennikov and Alexander Tsvetkov for their assistance in preparing the text.

 \end{document}